\documentclass[12pt]{article}
\usepackage{cite}
\usepackage{epsf}
\usepackage{graphicx}
\usepackage{psfrag}
\usepackage[dvips]{epsfig}
\usepackage[english]{babel}
\usepackage{amssymb,indentfirst}
\usepackage{amsmath}
\usepackage{verbatim}

\makeatletter
\renewcommand \thesection {\@arabic\c@section.}
\renewcommand\thesubsection   {\thesection\@arabic\c@subsection.}
\renewcommand\thesubsubsection{\thesubsection\@arabic\c@subsubsection.}
\makeatother

\def\starup#1{\mbox{$\raise1.8ex\hbox{$*$} \kern-.7em#1$}}
\def\krup#1{\mbox{$\raise1.8ex\hbox{$+$} \kern-1.0em#1$}}
\def\linup#1{\mbox{$\raise1.9ex\hbox{---} \kern-1.0em#1$}}

\begin{document}
\title{ 
Mass limits for the chiral color symmetry $G'$-boson from LHC dijet data   
}

\author{I.~V.~Frolov\footnote{{\bf e-mail}: pytnik89@mail.ru},
A.~D.~Smirnov\footnote{{\bf e-mail}: asmirnov@uniyar.ac.ru}
\\
{\small Division of Theoretical Physics, Department of Physics,}\\
{\small Yaroslavl State University, Sovietskaya 14,}\\
{\small 150000 Yaroslavl, Russia.}}
\date{}
\maketitle
\begin{abstract}
\noindent

The contributions of $G'$-boson predicted by the chiral color symmetry of quarks 
to the differential dijet cross-sections in $pp$-collisions at the LHC
are calculated and analysed in dependence on two free parameters of the model, 
the $G'$ mass $m_{G'}$ and mixing angle $\theta_G$.   
The exclusion and consistency $m_{G'}-\theta_G$ regions imposed by 
the ATLAS and CMS data on dijet cross-sections are found.   
Using the CT10 (MSTW~2008) PDF set 
we show that the $G'$-boson for $\theta_G=45^{\circ}$, i.e. the axigluon, 
with the masses 
 $m_{G'} < 2.3 \,\, (2.6) \,\, \mbox{TeV}$ and 
$m_{G'} < 3.35 \,\, (3.25) \,\, \mbox{TeV}$ 
is excluded at the probability level of $95\%$  by the ATLAS and CMS dijet data 
respectively. 
For the other values of  $\theta_G$  the exclusion limits are more stringent. 
The  $m_{G'}-\theta_G$ regions consistent with 
these data at  $CL=68\%$ and  $CL=90\%$ are also found.

\vspace{5mm}

\noindent

\end{abstract}

The search for the possible effects of new physics beyond the Standard Model (SM) 
is now one of the main goals of the experiments at the LHC.
By now there are many models predicting new particles which
can manifest themselves at the LHC energy through the possible new physics effects.   
Among such particles there are new quarks and leptons of the fourth generation of fermions, 
supersymmetric particles of the supersymmetry models, 
new scalar particles of two Higgs models, 
new gauge bosons of the weak left-right symmetry model, 
gauge and scalar leptoquarks of the four color quark-lepton symmetry models, etc.      
The unobservation of the new physics effects induced by these particles 
at the LHC will set new limits on the parameter of corresponding models.

One of such models which also predict new particles and can induce 
the new physics effects at the LHC energy is based on the gauge group 
of the chiral color symmetry of quarks 
\begin{eqnarray}
\label{chiral_group}
G_c = SU_{cL}(3)\!\times \! SU_{cR}(3)  \stackrel{\,M_{chc}\,}{\longrightarrow}  SU_c(3)
\end{eqnarray}            
which is assumed to be valid at high energies and is spontaneously broken 
to usual QCD $SU_c(3)$ at some low energy scale $M_{chc}$. 
The idea of the originally chiral character of $SU_c(3)$ color symmetry of quarks 
was firstly proposed and realized for particular case of $g_L=g_R$ 
in refs.~\cite{Pati:1975ze,Hall:1985wz,Frampton:1987ut,Frampton:1987dn} 
and then it was extended to the more general case of  
$g_L\neq g_R$~\cite{Cuypers:1990hb,Martynov:2009en,Martynov:2010ed,Martynov:2012zz}.   

The chiral color symmetry of quarks in addition to the usual gluons predicts 
immediately a new massive gauge particle -- the axigluon \; $G^A$ 
(in the case of $g_L=g_R$) with pure axial vector couplings to quarks  
or $G'$-boson (in the case of  $g_L\neq g_R$) with vector and axial
vector couplings to quarks defined by the gauge group~\eqref{chiral_group}.
In both cases this new particle interacts with quarks with coupling constants 
of order~$g_{st}$ and can induce appreciable effects in the processes with quarks.     
In particular $G'$-boson could give rise at the LHC to 
the charge asymmetry of $t \bar t$ production as well as to the appearance of 
a resonant peak in the invariant mass spectrum of dijet events.  
The possible effect of  $G'$-boson on the charge asymmetry of $t \bar t$ production 
at the LHC (and at the Tevatron) and the corresponding $G'$-boson mass limits 
have been discussed 
in~\cite{Martynov:2009en,Martynov:2010ed,Martynov:2012zz,Frolov:2013kpa}.

In the present paper we calculate and analyse the possible $G'$-boson contributions  
to the differential dijet cross-sections in $pp$-collisions at the LHC 
in comparison with the ATLAS~\cite{Aad:2013tea} 
and CMS~\cite{Khachatryan:2015sja, Buckley:2010jn}  
data on dijet cross-sections and find the $G'$-boson mass limits imposed 
by these experimental data. 
   
The basic gauge fields of the group~\eqref{chiral_group} $G^L_\mu$ and $G^R_\mu$  
form the usual gluon field $G_\mu$ and the field $G'_\mu$ of 
an additional $G'$-boson as superpositions 
%
\begin{align}
\label{GmuGnu}
G_\mu&= \frac{g_R G^L_\mu + g_L G^R_\mu}{\sqrt{(g_L)^2+(g_R)^2}} \equiv s_G \, G^L_\mu 
+ c_G \, G^R_\mu,\\
\label{G1muGnu}
G'_\mu&= \frac{g_L G^L_\mu -  g_R G^R_\mu}{\sqrt{(g_L)^2+(g_R)^2}} \equiv c_G \, G^L_\mu 
- s_G \, G^R_\mu,
\end{align}
%
where $G^{L,R}_{\mu} = G^{L,R}_{i\mu} t_i$, $G_{\mu} = G^i_{\mu} t_i$, 
$G'_{\mu} = G'^i_{\mu} t_i$,
 $i=1,2,...,8$, $t_i$ are the generators of $SU_c(3)$ group,
$s_G =\sin\theta_G, \, c_G =\cos\theta_G $, $\theta_G$ is $G^{L} - G^{R}$ mixing angle,
$tg\,\theta_G=g_R/g_L$ and $g_L, g_R$ are the coupling constants 
of the group~\eqref{chiral_group}.

The interaction of the $G'$-boson with quarks can be written as
\begin{equation}
\label{L1Gqq}
\mathcal{L}_{G'qq} =
g_{st}(M_{chc}) \, \bar{q} \gamma^\mu (v + a \gamma_5) G'_\mu q , 
\end{equation}
where 
\begin{eqnarray}
g_{st}(M_{chc}) = \frac{g_L g_R}{\sqrt{(g_L)^2+(g_R)^2}}
\nonumber
\end{eqnarray}
is the strong interaction coupling constant
at the mass scale $M_{chc}$ of the chiral color symmetry breaking 
and the vector and axial-vector coupling constants $v$ and $a$  
are defined by the group~\eqref{chiral_group} and with account 
of the relations~\eqref{GmuGnu},~\eqref{G1muGnu} take the form   
\begin{eqnarray}
v = \frac{c_G^2-s_G^2}{2 s_G c_G} = \cot(2\theta_G), \,\,\,\,
a = \frac{1}{2 s_G c_G} = 1 / \sin(2\theta_G). 
\label{eg:va}
\end{eqnarray}

After spontaneously breaking the symmetry~\eqref{chiral_group} the $G'$-boson 
acquires some mass $m_{G'}$ and as a result we obtain two free parameters in the model, 
the $G'$-boson mass $m_{G'}$ and the $G^{L} - G^{R}$ mixing angle $\theta_G$. 
For $g_L = g_R$ we have $\theta_G=45^\circ$, $v = 0, a =1$ 
and $G'$-boson in this case coincides with the axigluon.

The interaction~\eqref{L1Gqq} gives to the $G'$-boson the hadronic width    
\begin{eqnarray}
\Gamma_{G'} = \sum_{Q} \Gamma (G' \to Q\overline{Q}) , 
\label{GammaG'}
\end{eqnarray}
where
\begin{eqnarray}
\Gamma (G' \to Q\overline{Q}) =
\frac{\alpha_{s}\, m_{G'}}{6}
\Bigg[ \, v^2 \left(1+\frac{2m_Q^{2}}{m_{G'}^{2}}\right)
+ a^2 \left(1-\frac{4m_Q^{2}}{m_{G'}^{2}}\right) \Bigg] \sqrt{1-\frac{4m_Q^{2}}{m_{G'}^{2}}}
\label{GammaG'QQ}
\end{eqnarray}
is the width of $G'$-boson decay into $Q\overline{Q}$-pair.
From~\eqref{eg:va}--\eqref{GammaG'QQ} 
follow the next estimations for the relative width of $G'$-boson 
$\Gamma_{G'}/m_{G'}=0.11, \; 0.18, \; 0.41, \; 0.75, \; 1.71$
for  $ \theta_G=45^\circ, \; 30^\circ, \; 20^\circ, \; 15^\circ, \; 10^\circ $
respectively.

The chiral color symmetry of quarks needs  
the extensions of the fermion sector of the model  
for cancellations of the chiral $\gamma_5$-anomalies which are produced by quarks 
in the case of the chiral color symmetry~\eqref{chiral_group} 
as well as of the scalar one for giving necessary masses to the fermions 
and gauge particles of the model.  
For discussing these extensions the chiral color symmetry~\eqref{chiral_group} 
should be unified with the electroweak $SU_L(2) \! \times \! U(1) $ symmetry 
of the SM for example in the simplest way by the group  
\begin{eqnarray}
\label{G3321}
G_{3321} = SU_{cL}(3)\!\times \! SU_{cR}(3)\!\times \! SU_L(2)\!\times \! U(1),  
\end{eqnarray}
where the first two factors are given by the group~\eqref{chiral_group} 
and the second ones form the usual SM electroweak symmetry group.   

There are many approaches for cancellation of the chiral $\gamma_5$-anomalies 
produced by the SM fermions through extensions of the SM fermion sector 
by introducing additional exotic fermions.  
The most simple and natural looks the variant~\cite{Geng:prd1989} in which in the case 
of the group~\eqref{G3321} for each SM doublet $q$ of quarks  $q_a, \, a=1, 2$  
transforming under the group~\eqref{G3321} in the usual way         
\begin{eqnarray}
\label{transqL}
q^L: \hspace{10mm} (3, \, 1, \, 2, \, Y_{q^L}) , \\
\label{transqR}
q^R_a: \hspace{10mm} (1, \, 3, \, 1, \, Y_{q^R_a}) 
\end{eqnarray}
with the SM quark hypercharges  $Y_{q^L}=1/3, \, Y_{q^R_1}=4/3, \, Y_{q^R_2}=-2/3$ 
one needs in each generation the existence of two exotic quarks $\tilde {q}_a$ 
transforming under the group~\eqref{G3321} as        
\begin{eqnarray}
\label{transqLt}
\tilde{q}^L_a: \hspace{10mm} (1, \, 3, \, 1, \, Y_{\tilde{q}_a}) , \\
\label{transqRt}
\tilde{q}^R_a: \hspace{10mm} (3, \, 1, \, 1, \, Y_{\tilde{q}_a}) ,   
\end{eqnarray}
where $Y_{\tilde{q}_1}=\tilde{Y}, \, Y_{\tilde{q}_2}= - \tilde{Y} + 2/3$ 
are the hypercharges of the exotic quarks, $\tilde{Y}$ is an arbitrary hypercharge.     
With patricular choice of $\tilde{Y}=4/3$ the electric charges of the exotic quarks 
become the same as those of the SM quarks.   

Using the transformation laws~\eqref{transqLt},~\eqref{transqRt} 
and the relations~\eqref{GmuGnu},~\eqref{G1muGnu} one can obtain  
the interaction of the exotic quarks with the  $G'$-boson in the form   
\begin{equation}
\label{L1Gqtqt}
\mathcal{L}_{G'\tilde{q}\tilde{q}} =
g_{st}(M_{chc}) \, \bar{\tilde{q}} \gamma^\mu (v - a \gamma_5) G'_\mu \tilde{q} , 
\end{equation}
where the constants $v$ and $a$ are given by the expressions~\eqref{eg:va}. 
Additionally the exotic quarks have the vector-like interactions with the photon 
and SM $Z$-boson.  
As seen from~\eqref{L1Gqtqt} the axial-vector coupling constant 
of the exotic quarks with the $G'$-boson has the opposite sign relatively to that of 
the SM quarks, which ensures the cancellation of the chiral $\gamma_5$-anomalies 
in diagrams with the $G'$-bosons.

For giving necessary masses to the fermions and gauge particles 
the scalar sector of the model should be appropriately extended. 
The masses to the usual quarks and leptons can be given by 
the scalar doublets $(\Phi^{(1,2)}_{a})_{\alpha \beta}$ and $\Phi^{(3)}_{a}$   
transforming under the group~\eqref{G3321} as 
\begin{eqnarray}
\label{transPhi1}
(\Phi^{(1)}_{a})_{\alpha \beta} : \hspace{10mm} (3, \, \bar{3}, \, 2, \,-1) , \\
\label{transPhi2}
(\Phi^{(2)}_{a})_{\alpha \beta} : \hspace{10mm} (3, \, \bar{3}, \, 2, \,+1) , \\ 
\label{transPhi3}
\Phi^{(3)}_{a} : \hspace{10mm} (1, \, 1, \, 2, \, +1)  
\end{eqnarray}
with VEVs
$\langle (\Phi^{(b)}_{a})_{\alpha \beta} \rangle =
\delta_{\alpha \beta} \,\delta_{ab} \,\eta_{b}/(2\sqrt{3})$ for $b=1, 2$  
and 
$\langle \Phi^{(3)}_{a} \rangle = \delta_{a2} \,\eta_{3}/\sqrt{2}$, 
$a=1, 2$ is the $SU_L(2)$ index 
and $\alpha, \beta = 1,2,3$ are the $SU_{cL}(3)$ and $SU_{cR}(3)$ indices.  

The mass to the $G'$-boson can be given by the scalar field $\Phi^{(0)}_{\alpha \beta}$ 
transforming under the group~\eqref{G3321} as
\begin{eqnarray}
\label{transPhi0}
(\Phi^{(0)})_{\alpha \beta} : \hspace{10mm} (3, \, \bar{3}, \, 1, \, 0) 
\end{eqnarray}
with VEV  
$\langle \Phi^{(0)}_{\alpha \beta} \rangle = 
 \delta_{\alpha \beta} \, \eta_0 /(2 \sqrt{3}) $. 
After such symmetry breaking the $G'$-boson acquires the mass
\begin{equation}
m_{G'} = \frac{g_{st}}{s_G c_G} \, 
\frac{\sqrt{\eta_{1}^2 + \eta_{2}^2  + \eta_{0}^2}}{\sqrt{6}} .
\label{eq:mG1}
\end{equation}

The field $\Phi^{(0)}_{\alpha \beta}$ can interact with the exotic quarks as 
\begin{equation}
\label{LPhi0qtqt}
\mathcal{L}_{\Phi^{(0)}\tilde{q}\tilde{q}} =
\bar{\tilde{q}}^R_{i a \alpha}\,(h_a)_{ij}\,(\Phi^{(0)})_{\alpha \beta}\,
\tilde{q}^L_{j a \beta}
 + \bar{\tilde{q}}^L_{i a \alpha}\,(h_a^+)_{ij}\,((\Phi^{(0)})^+)_{\alpha \beta}\,
\tilde{q}^R_{j a \beta}, 
\end{equation}
where $(h_a)_{ij}$ form the matrices of Yukawa coupling constants, 
$i ,j$ are the generation indeces. 
By biunitary transformations these matrices can be reduced to the diagonal 
form $(h_{ia}) \delta_{ij}$ and after the symmetry breaking the mass eigenstates 
$\tilde{q}^{\,\prime}_{i a}$ of the exotic quarks pick up the masses        
\begin{equation}
m_{\tilde{q}^{\,\prime}_{i a}} = h_{i a} \frac{\eta_{0}}{2 \sqrt{3}} \, . 
\label{eq:mqt1ia}
\end{equation}

As a result of the decomposition $(3_L, \bar{3}_R) = 1_{SU_c(3)} + 8_{SU_c(3)}$  
of the representation $(3_L, \bar{3}_R)$ of the group~\eqref{chiral_group}   
the multiplets $(\Phi^{(1,2)}_{a})_{\alpha \beta}$ and $\Phi^{(0)}_{\alpha \beta}$
after the chiral color symmetry breaking give the color singlets
$(\Phi^{(1,2;0)}_{a})_{\alpha \beta} = 
\Phi^{(1,2)}_{0a} \, \delta_{\alpha \beta} / \sqrt{6} $ and 
$\Phi^{(0;0)}_{\alpha \beta} = \Phi^{(0)}_{0} \, \delta_{\alpha \beta} / \sqrt{6} $ 
as well as the $SU_c(3)$ octets
$(\Phi^{(1,2;8)}_{a})_{\alpha \beta} = \Phi^{(1,2)}_{ia} (t_i)_{\alpha \beta} $, 
$\Phi^{(0;8)}_{\alpha \beta} = \Phi^{(0)}_{i} (t_i)_{\alpha \beta} $. 
The color singlets $\Phi^{(1,2)}_{0a}$ and $\Phi^{(3)}_{a}$ are 
the $SU_L(2)$ doublets. These doublets are mixed and form the SM Higgs 
doublet $\Phi^{(SM)}_{a}$ with the SM VEV 
$\eta_{SM}  = \sqrt{\eta_{1}^2 + \eta_{2}^2  + \eta_{3}^2} \approx 250 \, GeV $ 
and two additional doublets $\Phi'_{a}$, $\Phi''_{a}$.   
As a result the chiral color symmetry reproduces in the scalar sector 
the SM Higgs doublet $\Phi^{(SM)}_{a}$  
and predicts the new scalar fields:
two colorless doublets $\Phi'_{a}$ and $\Phi''_{a}$, two doublets of color
octets $\Phi^{(1,2)}_{ia}$,
the color octet $\Phi^{(0)}_{i}$ and the colorless $SU_L(2)$ singlet 
$\Phi^{(0)}_{0}$ 
with the VEV $\eta_0$. The interactions of the doublet $\Phi^{(SM)}_{a}$ 
with the SM quarks and leptons and with $W^{\pm}$ and  $Z$ bosons 
have the standard form.

The differential cross section of the process $p p \rightarrow c d + X$  of 
the inclusive production of two partons $c$ and $d$ in $p p $ collisions can be 
written 
in the usual way as 
\begin{equation}
  \begin{split}
    &d\sigma (p p \rightarrow c d +X) = 
\sum_{ab} \int \int f^{p}_a(x_1;\mu_F) f^{p}_b(x_2;\mu_F)
    d\sigma(ab\rightarrow cd) dx_1 dx_2, 
\label{dsigmappcd}
  \end{split}
\end{equation}
where $a=q_i, \bar{q_i}, g; \,\,\, b=q_j,\bar{q_j}, g$ are the initial partons, 
$i, j =1, 2, ...,5 $ are the quark (or antiquark) flavour indices, 
$f^{p}_a(x;\mu_F)$ are the parton distribution function (PDF) 
of quark, antiquark or gluon, 
$x_{1,2}$ are the momentum fractions of proton 
carried by initial partons, $\mu_F$ is the factorization scale and  
%
\begin{equation}
  d\sigma(ab\rightarrow cd 
) 
=
  (2\pi)^4 \delta^{(4)}(p_3+p_4-p_1-p_2)\frac{|M_{ab\rightarrow cd}|^2}{2 \hat{s}}
  \frac{d^3 p_3}{2E_3 (2\pi)^3}\frac{d^3 p_4}{2E_4 (2\pi)^3}
\label{dsigmaabcd}
\end{equation}
are the parton differential cross sections, $p_{1,2}$ and 
$p_{3,4}=\{E_{3,4},\vec p_{3,4}\}$ are the four momenta 
of the initial and final partons, $\hat{s}=(p_1+p_2)^2$.

The amplitudes of parton processes $ab \rightarrow cd$ can be presented as 
\begin{equation}
  \begin{split}
    &M_{ab\rightarrow cd}= M_{ab\rightarrow cd}^{SM}+M_{ab\rightarrow cd}^{G',LO},  
  \end{split}
\label{Mabcd}
\end{equation}
where  
\begin{equation}
  \begin{split}
 &M_{ab\rightarrow cd}^{SM}= M_{ab\rightarrow cd}^{SM,LO}+M_{ab\rightarrow cd}^{SM,NLO}
  \end{split}
\label{MSMabcd}
\end{equation}
are the SM amplitudes calculated up to next to leading order and the amplitudes 
$M_{ab\rightarrow cd}^{G',LO}$ 
give the $G'$-boson contributions in the leading order. 

\begin{figure}[!ht]
\centerline{
\epsfysize=0.6
\textwidth
\epsffile{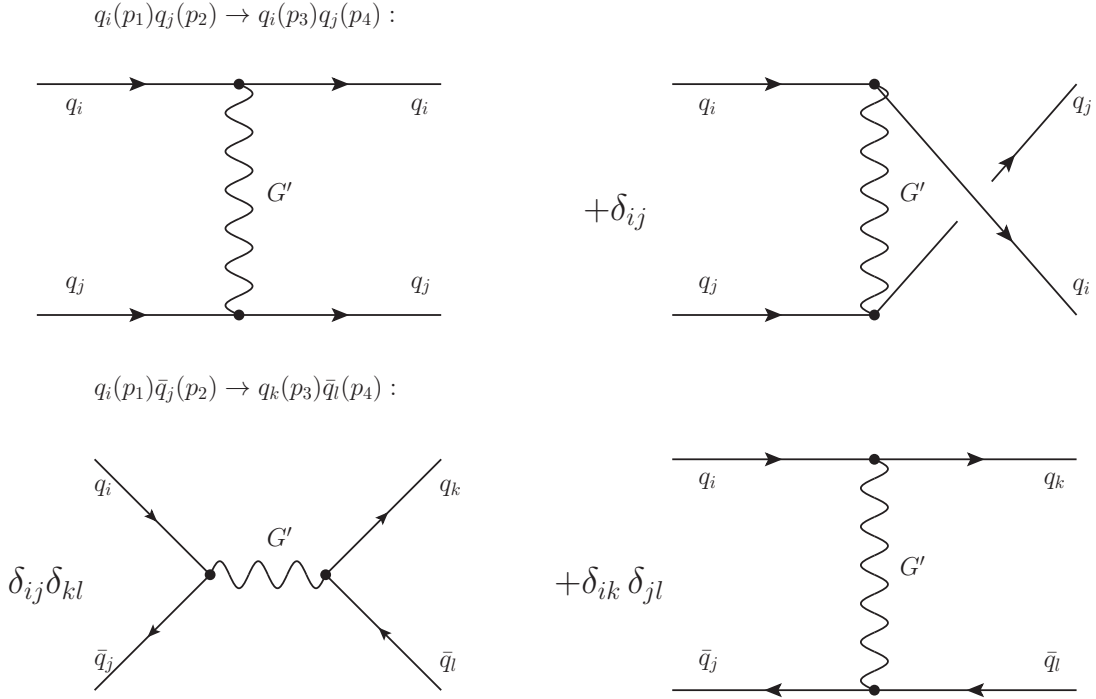}}
\caption{
Diagrams describing the leading-order $G'$-boson contributions to the amplitudes 
$M_{ab\rightarrow cd}^{G',LO}$ of the partonic processes  
$q_i q_j\rightarrow q_i q_j$  and $q_i \bar{q}_j \rightarrow q_k \bar{q}_l$   
(diagrams for the processes $\bar{q}_i \bar{q}_j\rightarrow \bar{q}_i \bar{q}_j$ 
can be obtained from the diagrams of the processes $q_i q_j\rightarrow q_i q_j$  
by changing the directions of the fermion lines)
}
\label{fig3}
\end{figure}

The amplitudes $M_{ab\rightarrow cd}^{G',LO}$ are described by the diagrams 
shown in the Fig.~\ref{fig3}. 
%

For the squared amplitudes $|M_{ab\rightarrow cd}|^2$ we use the expressions 
\begin{equation}
  \begin{split}
&|M_{ab\rightarrow cd}|^2= 
|M_{ab\rightarrow cd}^{SM}|^2+|M_{ab\rightarrow cd}^{G',LO}|^2 + 
2Re(M_{ab\rightarrow cd}^{SM,LO}\stackrel{*}{M}{}^{\!\!G',LO}_{\!\!ab\rightarrow cd}) + 
... , 
  \end{split}
\label{Mabcd2}
\end{equation}
where the dots denote the terms 
$2Re(M_{ab\rightarrow cd}^{SM,NLO}\stackrel{*}{M}{}^{\!\!G',LO}_{\!\!ab\rightarrow cd})$ 
omitted because of their smallness. 

For the calculation of the $G'$-boson contribution in dijet mass spectrum 
we use a package for calculation of Feynman diagrams and integration over 
multi-particle phase space CalcHEP~\cite{Belyaev:2012qa} 
and for calculation of differential dijet NLO cross section in the SM we use  
the program for computation of inclusive jet cross sections 
at hadron colliders MEKS~\cite{Gao:2012he}. 

As a result of calculations for the terms $|M_{ab\rightarrow cd}^{G',LO}|^2$ and 
$2Re(M_{ab\rightarrow cd}^{SM,LO}\stackrel{*}{M}{}^{\!\!G',LO}_{\!\!ab\rightarrow cd})$ 
in~\eqref{Mabcd2} we obtain the expressions   
\begin{eqnarray}
&& \hspace{-15mm} 
|M_{q_i q_j \rightarrow q_i q_j}^{G',LO}|^2 =
|M_{\bar{q}_i \bar{q}_j \rightarrow \bar{q}_i \bar{q}_j}^{G',LO}|^2= 
\nonumber
\\
&& \hspace{-15mm} 
\frac{4g^4_{st}(M_{chc})}{27}\left(\frac{3 \delta _{ij} 
\left(\hat{t}^2 A^2+\hat{s}^2 B\right)}{\left(\hat{u}-m_{G'}^2\right)^2}
-\frac{2 \delta_{ij}\hat{s}^2 B }{\left(\hat{t}-m_{G'}^2\right) 
\left(\hat{u}-m_{G'}^2\right)}\right.
\left.+\frac{3\left(\hat{u}^2 A^2+\hat{s}^2 B\right)}
{\left(\hat{t}-m_{G'}^2\right){}^2}\right), 
\label{MGGqiqj}
\\
&& \hspace{-15mm} 
|M_{q_i\bar{q}_j\rightarrow q_k\bar{q}_l}^{G',LO}|^2=
\frac{4g^4_{st}(M_{chc})}{27} 
\nonumber
\\
&& \hspace{-15mm} 
\left( \hspace{-1mm} \frac{3 \delta_{ik}\delta_{jl} 
\left(\hat{s}^2 A^2+\hat{u}^2 B\right)}
{\left(\hat{t}-m_{G'}^2\right){}^2} 
-\frac{2\delta _{ij}\delta _{jk} \hat{u}^2\left(\hat{s}-m_{G'}^2\right)  B }
{\left( \hspace{-1mm} \left(\hat{s}-m_{G'}^2\right)^2 \hspace{-1mm} + \hspace{-1mm} 
m_{G'}^2\Gamma_{G'}^2 \hspace{-1mm} \right) \hspace{-1mm} 
\left(\hat{t}-m_{G'}^2\right)}\right. 
\left.+\frac{3\delta_{ij}\delta_{kl} \left(\hat{t}^2 A^2+\hat{u}^2 B\right)}
{\left(\hat{s}-m_{G'}^2\right){}^2 \hspace{-1mm} + \hspace{-1mm} 
m_{G'}{}^2\Gamma_{G'}{}^2} \hspace{-1mm} \right) 
\label{MGGqiaqj}
\end{eqnarray}
and 
\begin{eqnarray}
&& \hspace{-15mm} 
2Re(M_{q_iq_j\rightarrow q_i q_j}^{SM,LO}\stackrel{*}
{M}{}^{\!\!G',LO}_{\!\!q_i q_j\rightarrow q_i q_j})=
2Re(M_{\bar{q}_i \bar{q}_j\rightarrow \bar{q}_i \bar{q}_j}^{SM,LO}\stackrel{*}
{M}{}^{\!\!G',LO}_{\!\!\bar{q}_i \bar{q}_j\rightarrow \bar{q}_i \bar{q}_j})= 
\nonumber
\\
&& \hspace{-15mm} 
\frac{8g^2_{st}(\mu_R)g^2_{st}(M_{chc})}{27}\left(\frac{3 \hat{u}^3 A+\hat{s}^2\left(3\hat{u}- 
\delta_{ij}\hat{t} \right)C}{\hat{t} \hat{u}\left(\hat{t} -m_{G'}^2\right)}\right.
\left.+\frac{\delta_{ij} \left(3 \hat{t}^3A+ \hat{s}^2 (3 \hat{t}-\hat{u})C\right)}
{\hat{t} \hat{u} \left(\hat{u}-m_{G'}^2\right)}\right), 
\label{MSMGqiqj}
\\
&& \hspace{-15mm} 
2Re(M_{q_i\bar{q}_j\rightarrow q_k\bar{q}_l}^{SM,LO}\stackrel{*}
{M}{}^{\!\!G',LO}_{\!\!q_i\bar{q}_j\rightarrow q_k\bar{q}_l})=
\frac{8g^2_{st}(\mu_R) g^2_{st}(M_{chc})}{27} 
\nonumber
\\
&& \hspace{-15mm}  
\left( \hspace{-1mm} \frac{\delta_{jl} \left(3 \delta_{ik} \hat{s}^3 A 
\hspace{-1mm} + \hspace{-1mm} \hat{u}^2 (3 \delta_{ik} \hat{s} 
\hspace{-1mm} - \hspace{-1mm} \delta_{ij} \delta_{jk} \hat{t})C\right)}
{\hat{s} \hat{t} \left(\hat{t} \hspace{-1mm} - \hspace{-1mm} m_{G'}^2\right)}\right. 
\left.+\frac{\left(\hat{s} \hspace{-1mm} - \hspace{-1mm} m_{G'}^2\right)
\left(\hat{u}^2 \left(3 \delta_{ij} \delta_{kl} \hat{t} \hspace{-1mm} - \hspace{-1mm}  
\delta_{ij} \delta_{jk}\hat{s} \right)C \hspace{-1mm} + \hspace{-1mm} 
3 \delta_{ij} \delta_{kl} \hat{t}^3 A\right)}
{\hat{s} \hat{t} \left(\left(\hat{s}-m_{G'}^2\right)^2+m_{G'}^2\Gamma _{G'}^2\right)}
 \hspace{-1mm} \right), 
\label{MSMGqiaqj}
\end{eqnarray}
where $A=\left(v^2-a^2\right), \,\, B=\left(a^4+6 a^2 v^2+v^4\right), 
\,\,  C=\left(v^2+a^2\right)$
and 
$\hat{s}=(p_1+p_2)^2$, \\
$\hat{t}=(p_1-p_3)^2$,  $\hat{u}=(p_2-p_3)^2$, 
$\mu_R$ is the renormalization scale.

In order to compare the cross-section defined by 
the equations~\eqref{dsigmappcd}--\eqref{MSMGqiaqj}
with the experimental dijet double-differential cross-sections measured by 
the ATLAS collaboration~\cite{Aad:2013tea} we have calculated 
the double-differential cross-sections averaged over 
the bins of ref.~\cite{Aad:2013tea} 
as  
\begin{eqnarray}
    &&  \hspace{-15mm}
\overline{\frac{d^2\sigma(pp\rightarrow jet \,\,jet)}{dm_{jj}d|y_*|}}=
\frac{1}{\Delta m_{jj}}\frac{1}{\Delta |y_*|} \,
\int \limits_{m_{jj}^{-}}^{m_{jj}^{+}} \,\,
\int \limits_{|y_*|^{-}}^{|y_*|^{+}}
\frac{d^2\sigma(pp\rightarrow jet \,\,jet)}{dm_{jj}d|y_*|} 
\,dm_{jj}\,d|y_*|, 
\label{dsigmappjjdmdmyav}
\end{eqnarray}
where 
$m_{jj}^{\pm} =  m_{jj} \pm \Delta m_{jj}/2$, \, 
$|y_*|^{\pm} =  |y_*| \pm \Delta |y_*|/2$, \,
$m_{jj}$ ($|y_*|$) and $\Delta m_{jj}$ ($\Delta |y_*|$) are the central value and 
the width of the invariant mass (half the rapidity separation) bin.     
The dijet double-differential cross-section 
$\frac{d^2\sigma(pp\rightarrow jet \,\,jet)}{dm_{jj}d|y_*|}$ in~\eqref{dsigmappjjdmdmyav} 
has been calculated from the cross-section~\eqref{dsigmappcd},~\eqref{dsigmaabcd} 
with $|M_{ab\rightarrow cd}|^2$ defined by 
the equations~\eqref{Mabcd2}--\eqref{MSMGqiaqj} 
by using the package CalcHEP~\cite{Belyaev:2012qa} 
and the program MEKS~\cite{Gao:2012he} with account the kinematic relations      
\begin{eqnarray}
&&    \hat{s} = x_1 x_2 s,     \hspace{5mm}
    \hat{t} = -\frac{\hat{s}}{2}(1-\tanh (y_*)),  \hspace{5mm}
    \hat{u} = -\frac{\hat{s}}{2}(1+\tanh (y_*)),  \hspace{5mm}
\label{stuxy}
\\ 
&&    y_{*}=\frac{y_3-y_4}{2}, \,\,\ 
y_{3,4}=\frac{1}{2}\ln\frac{E_{3,4}+p_{3,4;z}}{E_{3,4}-p_{3,4;z}} \, ,   
\label{y*y3y4}
\end{eqnarray}
where 
$\sqrt{s}$ is the center of mass energy of colliding protons, 
$y_{3,4}$ are the rapidities of the final partons and 
$m_{jj}=\hat{s}=x_1 x_2 s$ is the invariant mass of two jets.

For comparing the cross-section~\eqref{dsigmappcd}--\eqref{MSMGqiaqj} 
with the experimental dijet differential cross-sections measured by 
the CMS collaboration~\cite{Khachatryan:2015sja} we have calculated 
the differential cross-sections averaged over 
the bins of ref.~\cite{Khachatryan:2015sja} 
as  
\begin{eqnarray}
    &&  \hspace{-15mm}
\overline{\frac{d\sigma(pp\rightarrow jet \,\,jet)}{dm_{jj}}}=
\frac{1}{\Delta m_{jj}}
\int \limits_{m_{jj}^{-}}^{m_{jj}^{+}} \,\,
\int \limits_{0}^{|\eta_{*}|_{max}}
\frac{d^2\sigma(pp\rightarrow jet \,\,jet)}{dm_{jj} d|\eta_{*}|} \, 
d|\eta_{*}|,  
\label{dsigmappjjdmav}
\end{eqnarray}
where the dijet double-differential cross-section 
$\frac{d^2\sigma(pp\rightarrow jet \,\,jet)}{dm_{jj}d|\eta_*|}$ 
in~\eqref{dsigmappjjdmav} 
is obtained from
$\frac{d^2\sigma(pp\rightarrow jet \,\,jet)}{dm_{jj}d|y_*|}$ in~\eqref{dsigmappjjdmdmyav} 
by the substitution $y_*  \to \eta_* $ with  
\begin{eqnarray}
&&    \eta_{*}=\frac{\eta_3-\eta_4}{2}, \,\,\ 
\eta_{3,4}=-\ln[\tan(\theta_{3,4}/2)],      
\nonumber
\end{eqnarray}
where $\eta_{3,4}$ and $\theta_{3,4}$ are the pseudorapidities and 
the polar scattering angles of the final jets. 
The upper limit $|\eta_{*}|_{max}$ in~\eqref{dsigmappjjdmav} is defined 
by the CMS experiment. 
The variable $|\eta_{*}|$ relates to the pseudorapidity separation $|\Delta \eta_{jj}|$ 
used in ref.~\cite{Khachatryan:2015sja} as $|\Delta \eta_{jj}|=2|\eta_*|$.  

For the comparison of the experimental and theoretical results we use 
the variable~$\chi^2_r $  (,,reduced''~$\chi^2 $ ) defined as  
\begin{eqnarray}
\chi^2_r = \frac{1}{n} \sum_i^N \frac{(\sigma^{exp}_i - \sigma^{th}_i)^2}
{(\Delta\sigma^{exp}_i)^2},
\label{chikw}
\nonumber
\end{eqnarray}
where $\sigma^{exp}_i$ denote the experimental value of 
$\overline{\frac{d^2\sigma(pp\rightarrow jet \,\,jet)}{dm_{jj}d|y_*|}}$ 
in the case of the ATLAS data 
(or $\overline{\frac{d\sigma(pp\rightarrow jet \,\,jet)}{dm_{jj}}}$ 
in the case of the CMS data)
in the $i$-th bin,  
$\sigma^{th}_i$ is the corresponding theoretical value,   
$\Delta\sigma^{exp}_i$ is the experimental error of this value, 
$n=N-N_p$ is the number of degrees of freedom,
$N$ is the number of the bins under consideration and 
$N_p$ is the number of the free parameters of the model. 
In the further analysis we use the values $N_p=0$ for the SM 
and $N_p=2$ for the gauge chiral color symmetry model.

The ATLAS dijet double-differential cross-sections were measured~\cite{Aad:2013tea} 
for $pp$ collisions at $\sqrt{s}=7 \,\, \mbox{TeV}$ with $4.5 \,\, \mbox{fb}^{-1}$ 
as functions of the dijet mass $m_{jj}$ and 
half the rapidity separation $y^*=|y_3-y_4|/2$ of the two leading jets  
(the variable $y^*$ of ref.~\cite{Aad:2013tea} 
relates to our variable~\eqref{y*y3y4} as $y^*=|y_{*}|$).  
The measurements are performed in six ranges of $y^*$ in interval $0<y^*<3.0$  
in equal steps of $0.5$. The ATLAS collaboration provides the results 
as tables of measured dijet cross-section in $N=21$ dijet mass bins 
for all six ranges of $y^*$.

In order to compare our results with the ATLAS measurements 
we have calculated the double-differential 
cross-sections~\eqref{dsigmappjjdmdmyav} for the same dijet mass bins 
and the ranges of $y^*$ as the ATLAS ones and use in our calculations 
the ATLAS kinematic cuts on the final states 
\begin{eqnarray*}
  |y_i| &<& 3 \,\,\, (i=3,4), \\
  p_{T_{i}}&>& 100 \,\, \mbox{GeV}, 
\label{ATLAScuts}
\end{eqnarray*}
where $y_i$ and $p_{T_i}$ are the rapidity and transverse momentum of each jet. 
We set also the factorization ($\mu_F$) 
and renormalization ($\mu_{R}$) scales 
as $\mu = \mu_R=\mu_F=p_T^{max}e^{0.3 y^*}$ 
where  $p_T^{max}$ is the $p_T$ of the leading jet.  
In the ATLAS analysis~\cite{Aad:2013tea} the jets are clustered 
by the anti-$k_t$ algorithm using two values of the radius parameter, 
$R=0.4$ and $R=0.6$.  
For definiteness we perform our calculations and analysis of the ATLAS data 
with the value $R=~0.6$.     
To demonstrate the dependence of the results on the parton
distribution functions we use in our theoretical evaluations two PDF sets,  
CT10~\cite{Lai:2010vv} and MSTW~2008~\cite{Martin:2009iq}.  

We have calculated the double-differential 
cross-sections~\eqref{dsigmappjjdmdmyav} 
by means of the programs MEKS and CalcHEP with using the CT10 and MSTW~2008 PDF sets
with account of the contributions of the SM NLO QCD and $G'$-boson 
in each dijet mass bin and in all the ranges of $y^*$ 
for the model parameters $m_{G'}>1 \,\, \mbox{TeV}$ and 
$10^{\circ}<\theta_G<45^{\circ}$.   
The results accounting only the SM NLO QCD contribution are 
in good agreement with the cross-sections measured by the ATLAS 
(in all the bins the theoretical cross-sections consist with the experimental ones 
within the experimental errors). 
For example, for the range $0 \le y^* < 0.5$ and the jet radius parameter R = 0.6 
we obtain that $\chi^2_{r SM}=6.8/21=0.32$ in the case of using the CT10 PDF set  
and $\chi^2_{r SM}=15.5/21=0.74$ in the case of the MSTW~2008 PDF set.
%
\begin{figure}[!ht]
\centerline{
\epsfysize=0.8
\textwidth
\epsffile{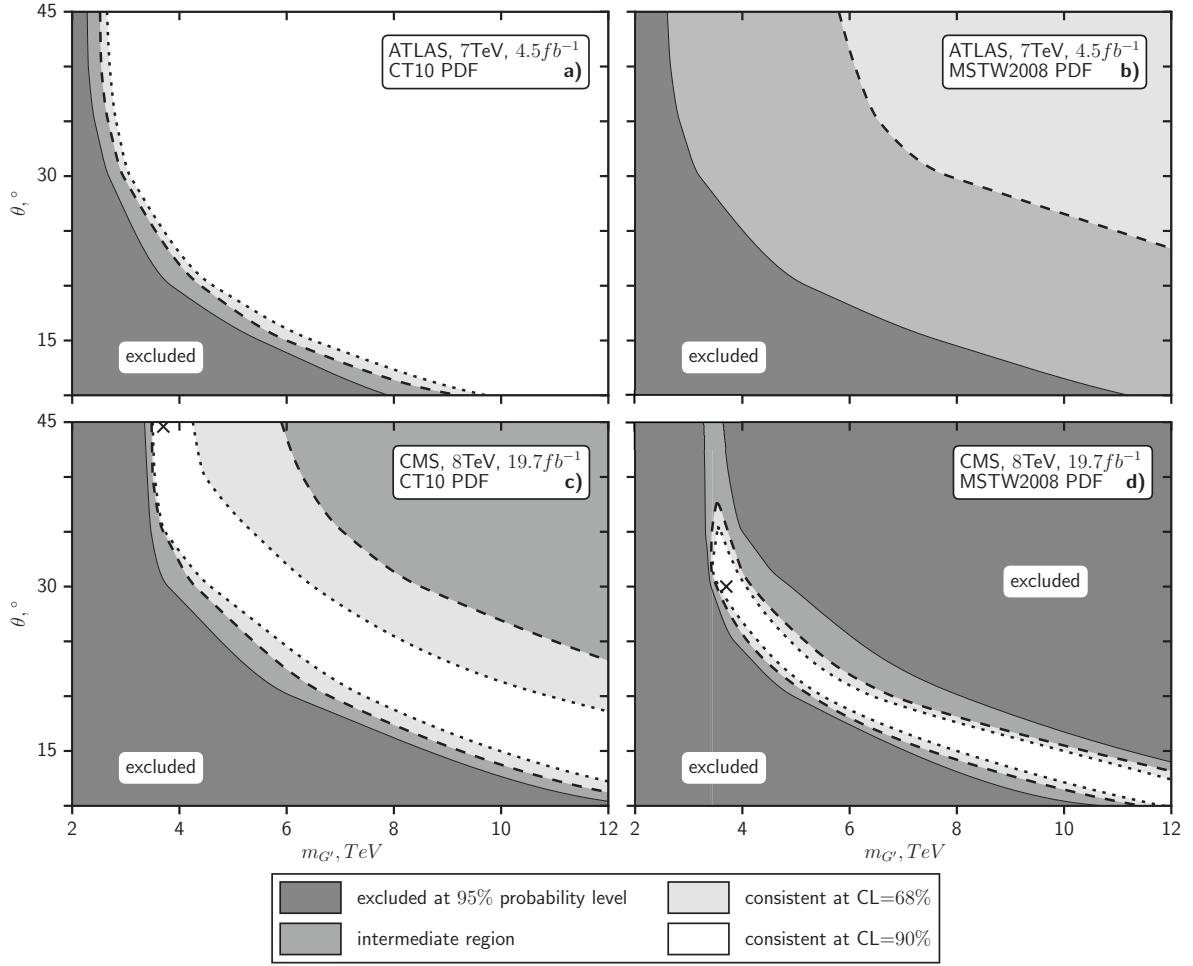}}
\caption{
The exclusion (at $95\%$ probability level) and consistency (at $CL=68\%$ and $CL=90\%$) 
$m_{G'}-\theta_G$ regions resulting from 
the ATLAS ($\sqrt{s}=7 \,\, \mbox{TeV}$ with $4.5 \,\, \mbox{fb}^{-1}$) 
and CMS ($\sqrt{s}=8 \,\, \mbox{TeV}$ with $19.7 \,\, \mbox{fb}^{-1}$) data 
on dijet cross sections with using the CT10 and MSTW~2008 PDF sets.    
}
\label{CMS_ATLAS_consistent}
\end{figure}

We have compared the results accounting simultaneously the contributions of  
the SM NLO QCD and $G'$-boson with the ATLAS data   
to find the regions of the parameters $m_{G'}$ and $\theta_G$ which are 
excluded by the ATLAS data at the appropriate level as well as the allowed ones.    

In the upper sections of the Fig.~\ref{CMS_ATLAS_consistent} we show the excluded 
and allowed $m_{G'}-\theta_G$ regions resulting from the ATLAS data on 
the double-differential dijet cross sections for the range $0 \le y^* < 0.5$ and 
the jet radius parameter R = 0.6 in the cases of using the CT10 (section~a)) 
and MSTW~2008 (section~b)) PDF sets. 
The solid line shows the bound of the~ $m_{G'}-\theta_G$ region excluded at 
the probability level of $95\%$ and the dashed and dotted lines show the bounds 
of $m_{G'}-\theta_G$ regions which are consistent with the data 
at $CL=68\%$ and  $CL=90\%$ respectively.  

As seen from the Fig.~\ref{CMS_ATLAS_consistent}~a),~b) 
in the case of using the CT10 (MSTW~2008) PDF set  
the $G'$-boson for $\theta_G=45^{\circ}$ (i.e. the axigluon)  
with the masses 
\begin{eqnarray}
	m_{G'}&<&2.3 \,\, (2.6) \,\, \mbox{TeV} %
\label{ATLASexclbounds}
\end{eqnarray}
is excluded at the probability level of $95\%$  by the ATLAS dijet data 
and for the other values of  $\theta_G$  the corresponding exclusion limits 
are more stringent.     
At the same time in dependence on  $\theta_G$  the $G'$-boson with masses 
\begin{eqnarray}
	m_{G'}&>&2.55 \,\, (5.8) \,\, \mbox{TeV} %
\label{ATLASconsbounds68}
\end{eqnarray}
is consistent with these data at $CL=68\%$  
and with masses  
\begin{eqnarray}
	m_{G'}&>&2.65 \,\, \mbox{TeV} %
\label{ATLASconsbounds90}
\end{eqnarray}
at $CL=90\%$ in the case of using the CT10 PDF set. 

The CMS search~\cite{Khachatryan:2015sja} for dijet resonances was done 
at $\sqrt{s}=8 \,\, \mbox{TeV}$ with $19.7 \,\, \mbox{fb}^{-1}$ of
the data and the dijet differential cross-section as function of the dijet mass 
was measured and is accessible at the HepData~\cite{Buckley:2010jn}.
For comparison with CMS data we used in our numerical calculations 
the CMS dijet search criteria 
\begin{eqnarray*}
	|\eta_i| &<&  2.5 \,\,\, (i=3,4),  \\
	|\Delta \eta_{jj}| &<& 1.3 ,  \\
	H_{T} &>& 650 \,\, \mbox{GeV}, 
\label{CMScriteria}
\end{eqnarray*}
where $\eta_i$ is the pseudorapidity of each jet, 
$|\Delta \eta_{jj}|=2|\eta_*|$ is the pseudorapidity separation 
of the two jets and $H_{T}$ is the scalar sum of the jet $p_T$,  
and the values of the renormalization and factorization scales 
$\mu = \mu_R=\mu_F=p_T^{max}$ and of the radius parameter $R=1.1$. 

By means of the programs MEKS and CalcHEP with using the CT10 and MSTW~2008 PDF sets
we have calculated the dijet differential cross-sections~\eqref{dsigmappjjdmav}   
with account of the contributions of the SM NLO QCD and $G'$-boson 
in each dijet mass bin for the model parameters $m_{G'}>1 \,\, \mbox{TeV}$ and 
$10^{\circ}<\theta_G<45^{\circ}$.    

The comparison of the theoretical results with the CMS dijet data occurs to be 
slightly more complicated then that in the case of the ATLAS data.  

\begin{figure}[!ht]
\centerline{
\epsfysize=0.57
\textwidth
\epsffile{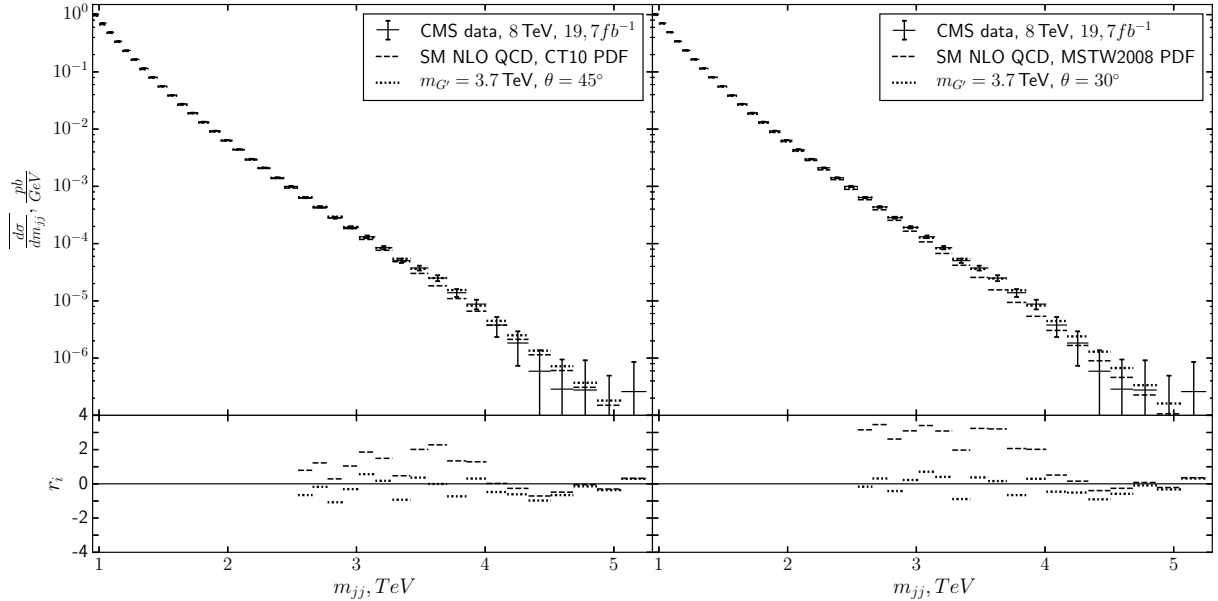}}
\caption{
The dijet differential cross-section 
$\overline{\frac{d\sigma(pp\rightarrow jet \,\,jet)}{dm_{jj}}}$ 
calculated using the CT10 (MSTW2008) PDF set 
for the jet radius parameter R = 1.1
with account of the contributions of the SM NLO QCD and  $G'$-boson 
with $m_{G'}=3.7 \,\, \mbox{TeV}$,  $\theta_G=45^\circ$    
($m_{G'}=3.7 \,\, \mbox{TeV}$,  $\theta_G=30^\circ$)    
 in comparison with the CMS dijet data.  
}
\label{CMS_CT10_MSTW}
\end{figure}

In the upper left (right) section of the Fig.~\ref{CMS_CT10_MSTW} we show  
the CMS data~\cite{Khachatryan:2015sja,Buckley:2010jn} 
with their experimental errors by the solid lines
and the dijet differential cross-section~\eqref{dsigmappjjdmav}   
calculated with accounting the SM NLO QCD contribution  
for the case of using the CT10 (MSTW2008) PDF sets by the dashed lines, 
for the jet radius parameter $R~=~1.1$. 
In the corresponding lower sections of the Fig.~\ref{CMS_CT10_MSTW} we show the relations
$r_i = (\sigma^{exp}_i - \sigma^{th}_i)/\Delta\sigma^{exp}_i$,  
where $\sigma^{exp}_i$  ($\sigma^{th}_i$) denote the experimental (theoretical) value 
of $\overline{\frac{d\sigma(pp\rightarrow jet \,\,jet)}{dm_{jj}}}$ 
in the $i$-th bin and $\Delta\sigma^{exp}_i$ is the experimental error of this value, 
the dashed lines correspond also to the case of accounting 
only the SM NLO QCD contribution to $\sigma^{th}_i$.  

As seen from the Fig.~\ref{CMS_CT10_MSTW} in the bins from the region 
$2.2 \,\, \mbox{TeV}<m_{jj}<4.0 \,\, \mbox{TeV}$ the experimental values 
of the dijet differential cross-section exceed those 
calculated with accounting the SM NLO QCD contribution by more or about 
one experimental errors. 

For the further analysis we use the last $N=22$ bins of the Fig.~\ref{CMS_CT10_MSTW} 
from the range $m_{jj}>2.0 \,\, \mbox{TeV}$.  
In this range the results of the SM NLO QCD calculations agree with the CMS dijet data 
with $\chi^2_{r SM} = 23.4/22=1.1$ ($\chi^2_{r SM} = 106.7/22=4.8$) 
in the case of using the CT10 (MSTW2008) PDF set. 
As seen, this agreement is not sufficiently good, especially in the case 
of using the MSTW2008 PDF set.     

In the case of the simultaneous account of the contributions of the SM NLO QCD 
and $G'$-boson we vary two free parameters of the gauge chiral color symmetry model 
($m_{G'}$ and $\theta_G$) to minimize $\chi^2_r$. 
It is found that in the case of using the CT10 (MSTW2008) PDF set 
the minimum of $\chi^2_r$ is at the values     
\begin{eqnarray}
&&	m_{G'} = 3.7 \,\, (3.7) \,\,  \mbox{TeV}, \,\,\,\, 
\theta_G = 45^\circ \,\, (30^\circ).  
\label{mG1thetaGCMSmin}
\end{eqnarray}

The dijet differential cross-section~\eqref{dsigmappjjdmav} 
calculated with simultaneous accounting the contributions of the SM NLO QCD 
and $G'$-boson with parameters~\eqref{mG1thetaGCMSmin} 
and the corresponding relations $r_i$ are shown in the Fig.~\ref{CMS_CT10_MSTW} 
by the dotted lines. This cross-section 
agrees with the corresponding CMS data (solid lines) with 
$\chi^2_{r\, min}=7.8/20=0.39 \,\,  (7.2/20=0.36)$  
when using the CT10 (MSTW2008) PDF set,   
i.e. better in comparison with the quoted above 
$\chi^2_{r SM} = 1.1 \,\, (4.8)$ 
in the case of accounting only the SM NLO QCD contribution (dashed lines). 

We have found and analysed the exclusion and consistency 
$m_{G'}-\theta_G$ regions resulting from the CMS dijet data.   
In the lower sections of the Fig.~\ref{CMS_ATLAS_consistent} we show the excluded 
and allowed $m_{G'}-\theta_G$ regions resulting from the CMS data on 
the differential dijet cross sections for R = 1.1 
in the cases of using the CT10 (section~c)) and MSTW~2008 (section~d)) PDF sets. 
In these sections the crosses refer to the points~\eqref{mG1thetaGCMSmin} 
and the other notations are the same as those in the upper ones.   

As seen from the Fig.~\ref{CMS_ATLAS_consistent}~c),~d) 
in the case of using the CT10 (MSTW~2008) PDF set  
the $G'$-boson for $\theta_G=45^{\circ}$ (i.e. the axigluon)  
with the masses 
\begin{eqnarray}
	m_{G'}&<&3.35 \,\, (3.25) \,\, \mbox{TeV} %
\label{CMSexclbounds}
\end{eqnarray}
is excluded at the probability level of $95\%$  by the CMS dijet data 
and for the other values of  $\theta_G$  the corresponding exclusion limits 
are more stringent. 
In the up-right part of the Fig.~\ref{CMS_ATLAS_consistent}~d) 
there is an additional exclusion region but this region is caused mainly by 
the mentioned above not sufficiently good agreement between the CMS dijet data and            
the SM NLO QCD calculations with using the MSTW2008 PDF set. 
It is also seen that in the both cases c) and d) there are the  $m_{G'}-\theta_G$ regions 
which are consistent with the CMS dijet data at  $CL=68\%$ and  $CL=90\%$.      

It should be noted that our results are obtained with account only the $G'$-boson 
contribution to the dijet cross-sections and generally speaking the presence 
of the new particle (the exotic quarks an additional scalar particles) in the model 
could influence on these results.      

Through their OCD interaction with gluons and interaction~\eqref{L1Gqtqt} 
with $G'$-boson the exotic quarks can contribute to the dijet cross-sections 
as well as give the analogous to~\eqref{GammaG'},~\eqref{GammaG'QQ} additional 
contributions to the $G'$-boson hadronic width.    
We have calculated the contributions of the exotic quarks to the dijet cross-sections 
analogously to the case of the usual quarks  
and found that the CMS data~\cite{Khachatryan:2015sja, Buckley:2010jn} 
are consistent within experimental errors with the existence of  
the exotic quarks with masses 
$m_{\tilde{q}^{\,\prime}_{i a}} \gtrsim 900 \,\, \mbox{GeV}$. 
It is worthy to note that the current experimental low mass limits for additional 
heavy quarks are of about a few hundreds GeV. For example the ATLAS Collaboration 
excludes the $t^{\prime}$~quark with the mass 
$m_{t^{\prime}} < 790 \,\, \mbox{GeV} $~\cite{ATLAS-CONF-2013-018} 
and the CMS exclusion limit for $T_{5/3}$ is of 
$m_{T_{5/3}} < 800 \,\, \mbox{GeV} $~\cite{Chatrchyan:prl2014}.    

Accounting in~\eqref{GammaG'},~\eqref{GammaG'QQ} also the contributions 
of three generations of the exotic quarks with   
$m_{\tilde{q}^{\,\prime}_{i a}} = 900 \,\, \mbox{GeV}$ 
we obtain that the $G'$-boson peak becomes more wide and low, which slightly 
effects the results. For example instead of~\eqref{CMSexclbounds} we obtain 
in this case the mass limit 
$m_{G'} < 3.25 \,\, (3.23) \,\, \mbox{TeV}$, 
as seen the deviations are 
 $\Delta m_{G'}=-100 \,\, (-20) \,\, \mbox{GeV}$.        
Because of~\eqref{eq:mqt1ia},~\eqref{eq:mG1} 
the exotic quarks can be more heavy. For 
$m_{\tilde{q}^{\,\prime}_{i a}} = 1.5 \,\, \mbox{TeV}$ 
the deviations are small  
 $\Delta m_{G'} \approx -(5 - 20) \,\, \mbox{GeV}$  
and for $m_{\tilde{q}^{\,\prime}_{i a}} > m_{G'}/2 $  
the effect of the exotic quarks on the the dijet cross-sections 
near the $G'$-boson peak becomes negligible.

As mentioned above the model under consideration reproduces in the scalar sector 
the SM Higgs doublet $\Phi^{(SM)}_{a}$ and  
predicts new colorless doublets $\Phi'_{a}$ and $\Phi''_{a}$ and two doublets of color
octets $\Phi^{(1,2)}_{ia}$. 
Although the SM Higgs doublet $\Phi^{(SM)}_{a}$ interacts with 
the SM fermions and $W^{\pm}$ and  $Z$ bosons in the standard way 
the new doublets could contribute to the signal-strengths of the Higgs decays 
through the loop corrections. These loop contributions depend on the details of 
the interactions of the Higgs doublet with the new scalar doublets 
(coupling constants, mixings, masses).      
At the present time the signal-strengths of the partial Higgs decays are measured 
with accuracy of about 20 - 30~\%~\cite{ATLAS_EPJ_C_2016}. 
For example the signal-strength is equal to   
$\mu_{H\to \gamma \gamma} = 1.17 \pm 0.27$  
for the decay $H\to \gamma \gamma$  and has the global value   
$\mu = 1.18^{+0.15}_{-0.14}$.    
All the measured signal-strengths are compatible with the SM predictions.   
It seems that the uncertainty in the Higgs doublet interactions with 
the new scalar doublets allows at the present time  to satisfy 
the current experimental values of the signal-strengths of the Higgs decays.

In conclusion, we summarize the results of this paper. 

The possible  contributions of $G'$-boson predicted by the chiral color symmetry 
of quarks to the differential dijet cross-sections in $pp$-collisions at the LHC 
are calculated and analysed in dependence on two free parameters of the model,
the $G'$ mass $m_{G'}$ and mixing angle $\theta_G$. 

Using the ATLAS~\cite{Aad:2013tea} and CMS~\cite{Khachatryan:2015sja, Buckley:2010jn}  
data on dijet cross-sections we find the $G'$-boson mass limits 
(in dependence on $\theta_G$) imposed by these experimental data. 
In particular, it is found that 
in the case of using in theoretical calculations the CT10 (MSTW~2008) PDF set 
the $G'$-boson for $\theta_G=45^{\circ}$ (i.e. the axigluon)  
with the masses 
\begin{eqnarray}
\nonumber
	m_{G'}&<&2.3 \,\, (2.6) \,\, \mbox{TeV} \\ 
\nonumber
&& \hspace*{-75mm} \mbox{and}  \\ 
\nonumber
	m_{G'}&<&3.35 \,\, (3.25) \,\, \mbox{TeV} %
\end{eqnarray}
is excluded at the probability level of $95\%$  by the ATLAS and CMS dijet data 
respectively.  
For the other values of  $\theta_G$  the corresponding exclusion limits 
are more stringent. 
The  $m_{G'}-\theta_G$ regions which are consistent with the ATLAS and CMS dijet data 
at  $CL=68\%$ and  $CL=90\%$ are also found. 
The possible effect of new fermion 
and scalar particles of the model on these mass limits is briefly discussed.

\bibliographystyle{science}
\bibliography{frolov}

\end{document}